\documentclass[pre,twocolumn,showpacs,superscriptaddress,floatfix]{revtex4}
 \usepackage{graphicx}

 \newcommand{\la}{\lambda}
 \newcommand{\xv}{\vec{x}}
 \newcommand{\ep}{\epsilon}
 \newcommand{\tN}{\tilde{N}}
 \newcommand{\tP}{\tilde{P}}
 \newcommand{\nb}{\bar{N}(t)}
 \newcommand{\pb}{\bar{P}(m,t)}
 \begin{document}

 \title{Kang-Redner Anomaly in Cluster-Cluster Aggregation}
 \author{Supriya Krishnamurthy}
 \email{supriya@santafe.edu}
 \affiliation{Santa Fe Institute, 1399 Hyde Park Road, Santa Fe, NM 87501,
USA}
 \author{R. Rajesh}
 \email{r.ravindran1@physics.ox.ac.uk}
 \affiliation{Department of Physics - Theoretical Physics, University of
Oxford, 1 Keble Road, Oxford OX1 3NP, UK}
 \author{Oleg Zaboronski}
 \email{olegz@maths.warwick.ac.uk}
 \affiliation{Mathematics Institute, University of Warwick, Gibbet Hill
Road, Coventry CV4 7AL, UK}

 \date{\today}

 \begin{abstract}
 The large time, small mass, asymptotic behavior of the average mass
distribution $\pb$ is studied in a $d$-dimensional system of diffusing
aggregating particles for $1\leq d \leq 2$. By means of both a
renormalization group computation as well as a direct re-summation of
leading terms in the small reaction-rate expansion of the average mass
distribution, it is shown that $\pb \sim \frac{1}{t^d} \left(
\frac{m^{1/d}}{\sqrt{t}} \right)^{e_{KR}}$ for $m \ll t^{d/2}$, where
$e_{KR}=\epsilon +O(\epsilon ^2)$ and $\epsilon =2-d$.  
In two dimensions,
it is shown that $\pb \sim \frac{\ln(m)  \ln(t)}{t^2}$ for $ m \ll t/
\ln(t)$. Numerical simulations in two dimensions supporting the analytical
results are also presented. 
 \end{abstract}
 \maketitle

\section{\label{sec1}Introduction}

Reaction-diffusion systems in low dimensions provide an excellent testing
ground for developing our understanding of the fluctuation effects in
complex systems far from equilibrium. A great deal of information, both
numerical and analytical, has been gathered over the past twenty years to
show that the evolution of statistical properties in simple reaction
diffusion systems in low dimensions is anomalous, in the sense that it
does not follow the corresponding mean field equations (see \cite{Privman}
for a short review). In low enough dimensions, in the instances where an
exact solution is lacking, there are no formal methods by which the
exponents characterizing the different physical quantities may be
calculated.  The renormalization group method (see \cite{Cardy} for a
review) provides the only systematic way to calculate these exponents and
thus understand fluctuation-dominated kinetics in reaction-diffusion
systems. In this paper, we consider the model of irreversible aggregation
of diffusing, massive particles $A_{i}+A_{j} \stackrel{\la}{\rightarrow}
A_{i+j}$, in dimensions $1\leq d \leq 2$, and use the renormalization
group method to calculate the small mass ($m\ll t^{d/2}$) behavior of
$\bar{P}(m,t)$, the mean density of particles of mass $m$ at time $t$. As
we explain later in this section, this problem of determining  $\bar{P}(m,t)$
requires using
the full power of
the renormalization group method.
This is
unlike many other problems where just considering the rate equations 
with
a renomalised reaction rate $\lambda$ is enough to obtain the right
answer.

We now give a more precise definition of the cluster-cluster aggregation
(CCA) model $A_i+A_j \stackrel{\la}{\rightarrow} A_{i+j}$ and review
earlier relevant results. Consider a $d$-dimensional lattice and particles
that possess a positive mass. Given a configuration of particles on this
lattice, the system evolves in time via the following microscopic moves:
(i) With rate $D$, each particle hops to a nearest neighbor and (ii) with
rate $\la$, two particles on the same lattice site aggregate together to
form a new particle whose mass is the sum of masses of the two constituent
particles. As time increases, the number of particles decreases due to
collisions and ultimately when $t\rightarrow \infty$, all particles
coagulate together to form one massive aggregate. However, at finite
times there is a well defined average mass distribution $\bar{P}(m,t)$ 
which is of interest to determine. It will be shown later 
(see the text following Eq. (\ref{gr}))
that the large time limit of this model is the same 
as the large $\la$ limit. The $\lambda \rightarrow \infty$ limit
was studied numerically by Kang and Redner in one and
two dimensions \cite{KR}. It was shown that $\bar{P}(m,t)$ has the scaling
form $\bar{P}(m,t)= t^{-d} f(mt^{-d/2})$, where $d \leq 2$ is the dimension. 
The two exponents 
are easily determined from the two conditions $\int m \bar{P}
(m) dm \sim t^0$ (mass conservation) and $\int \bar{P}(m) dm \sim t^{-d/2}$ 
(recurrence of random
walks). The small mass behavior of $\bar{P}(m,t)$ can be obtained by knowing
the small $x$ behavior of the scaling function $f(x)$. On the basis of
numerical simulations it was conjectured in \cite{KR} that $f(x) \sim
x^{(2-d)/d}$. In one dimension, the model can be solved exactly
\cite{Spouge,satya}; it was shown that $f(x) \sim x$ or equivalently
$\bar{P}(m,t) \sim m t^{-3/2}$ for $m\ll \sqrt{t}$. The one dimensional
solution uses the property of ordering of particles on a line and is not
generalizable to higher dimensions. 
In two dimensions, $f(x)$ was seen numerically to increase
with $x$ for small $x$ \cite{KR}. 
Also, in two dimensions, the scaling function
could be determined in the limit of fixed $x$ for $t \rightarrow \infty$,  
where $x = m \ln(t)/t$. In this case
it was shown \cite{Oleg} that  $\pb = t^{-2} \ln^2(t) e^{-x}$  for
$x \ll \ln^{1/2}(t) $ and $|\ln (x) | \ll |\ln(t)|$.
This result however, becomes incorrect in the limit when $m$ is fixed 
as $t \rightarrow \infty$.

In this paper, we compute $\pb$ in $1\leq d \leq
2$ in the limit $t \rightarrow \infty, m/m_{0}=\mbox{fixed}$, where
$m_{0}$ is the mass of the lightest particle at $t=0$. 
We show that $\pb \sim
\frac{1}{t^d} \left( \frac{m^{1/d}}{\sqrt{t}} \right)^{e_{KR}}$ for $m
\ll
t^{d/2}$, where $e_{KR}=\epsilon +O(\epsilon ^2)$ and $\epsilon =2-d$.  In
two dimensions, it is shown that $\pb \sim \frac{\ln(m)  \ln(t)}{t^2}$ for
$m \ll t/ \ln(t)$.
These results provide a theoretical basis to the
results obtained by numerical methods in \cite{KR}. 

The CCA model may also be considered to be a special case of the more
general model in which the aggregation kernel is mass dependent. For a
review of results on the rate equation approach to this problem see
\cite{aldous,oshanin}. The dependence of $\pb$ on $m$ in one dimension in
this more general model has also been studied \cite{alava}. 
In this paper, we
will restrict ourselves to the aggregation kernel which is mass
independent;  {\it i.e.}, the rates $\la$ and $D$ are independent of mass.
When the mass is ignored and $\la \rightarrow \infty$, the CCA model
reduces to the well studied $A+A\rightarrow A$ model
\cite{avraham}.  The CCA model and its
variants also find application in a large number of physical systems including
colloidal suspensions \cite{White}, irreversible polymerisation
\cite{Fried}, aerosols and cloud formation \cite{Fried}, river networks
\cite{scheidegger} and coarsening phenomena \cite{coarsening}. 

Field theoretic methods have been previously used to study complex systems
far from equilibrium (see \cite{Cardy1} for a review).  We briefly review
results relevant to reaction-diffusion systems.  In some earlier
works \cite{Mik,Peliti}, the effective reaction rate and the decay
exponent of the average particle density were computed for the
$A+A\rightarrow A (\emptyset)$ model.  The renormalization group study of
the same model with sources was done in \cite{Droz}.  In \cite{Lee} the
systematic renormalization group procedure for the computation of average
density and density-density correlation function in $kA\rightarrow
\emptyset$ reaction was developed. In \cite{LeeCardy} renormalization
group analysis of $A+B\rightarrow \emptyset$ reaction in $d>2$ was used to
study the effects of initial fluctuations on the late time decay of
particle densities. The renormalization group technology developed by
Peliti, Lee and Cardy \cite{Peliti,Lee,LeeCardy} was used to compute the
average mass distribution of clusters in the CCA model
in the intermediate mass range in \cite{Oleg}. 

It turns out, however, that as far as the study of scaling properties of
one-point correlation functions in most reaction-diffusion systems is
concerned, renormalization group is not a vital tool. Consider, for
example, a single species annihilation model $A+A\rightarrow \emptyset$.
Once the renormalization of effective reaction rate is understood, the
correct density decay exponent can be obtained from simple dimensional
arguments \cite{Cardy1}. Alternatively, one can use simple random walk
arguments or use the Smoluchowski approximation 
(we refer to the case in which the reaction rate is
replaced by a time dependent reaction rate as the Smoluchowski approximation)
\cite{KR,Smolappr}, to obtain
the correct values of decay exponents. A renormalized mean field theory
or, alternatively, a version of Smoluchowski theory can also be used to
compute the average mass distribution in cluster-cluster aggregation for
intermediate masses \cite{Krapivsky}. 

In the CCA model considered in this paper, it turns out that the
stochastic field $P(m,\xv,t)$, describing the continuous limit of the local
mass distribution, has a non-zero anomalous dimension in $d<2$. The
scaling exponent governing the dependence of the average mass distribution
$\bar{P}(m,t)$ on mass is proportional to the anomalous dimension of the
operator corresponding to the local mass distribution. As explained
in Section ~\ref{sec2} any approximation
scheme which disregards this anomaly (such as the
Smoluchowski approximation) predicts that $\pb \sim m^{0}$
when $m \ll t^{d/2}$, for any dimension. 
This is in contradiction with both numerical
results \cite{KR} as well as the exact result in one dimension
\cite{Spouge,satya}.  The full power of renormalization group analysis has
thus to be brought to bear in order to compute the anomalous dimension of
$P(m,\xv,t)$ in the form of the $\epsilon = 2-d$ expansion. 
Since the peculiarity in the 
small mass distribution of cluster-cluster aggregation
is first discussed in \cite{KR}
we refer to this as the {\it Kang-Redner} anomaly. 

In addition to the calculation of $\pb$, we found many similarities
between the problem of cluster-cluster aggregation and the problem of weak
turbulence (see \cite{Zakh} for a review on weak turbulence). We elaborate
on this connection in Sec.~\ref{sec6}, where the Kang-Redner anomaly is
interpreted as a violation of the Kolmogorov (constant flux) spectrum of
particles in mass space due to strong flux fluctuations. 

The rest of the paper is organized as follows.  Section~\ref{sec2}
contains a discussion of the stochastic integro differential equation
satisfied by $P(m,\xv,t)$. The mean field results, as well as the reasons for
their failure in low dimensions are also included.  In Sec.~\ref{sec3}, we
analyze the large time asymptotic behavior of $\pb$ in $d < 2$ using the
renormalization group method. We do the same for
$d=2$ in Sec.~\ref{sec4}.
In Sec.~\ref{sec5}, we rederive the results of Sec.~\ref{sec3} 
and ~\ref{sec4} using
explicit re-summation of all diagrams giving the leading contribution to
the average mass density in the limit of large time. Reasons for the
failure of the Smoluchowski theory then become more transparent.  In
Sec.~\ref{sec6} we elaborate on the connections with weak turbulence. 
Finally, we conclude with a summary and discussion of open problems in
Sec.~\ref{sec7}.

\section{\label{sec2}Constant kernel cluster-cluster aggregation and mean
field analysis.}

The problem of computing density
correlation functions in $d$-dimensional stochastic processes  can be 
reformulated as an 
effective equilibrium
problem in $(d+1)$-dimensions with the help of Doi-Zeldovich-Ovchinnikov
trick \cite{Doi}. One can then attempt to solve the problem using the
powerful methods of statistical field theory, in particular those based on
renormalization group ideas. 
Starting from the lattice version of the CCA model, we would like to
derive the corresponding field theory and the Langevin equation obeyed by
$P(m,\xv,t)$, where $P(m,\xv,t) dm dV$ is the number of particles 
with masses in
the interval $[m, m+dm]$ in the volume $dV$.  It was shown in \cite{Oleg},
that the problem of finding all correlation functions of the local mass
distribution is equivalent to finding all moments of the following
stochastic integro-differential equation (stochastic Smoluchowski
equation): 
 \begin{eqnarray}
\lefteqn{
\left( \frac{\partial}{\partial t} - D \Delta \right) P (m, \xv, t) =
\la P*P} \quad \nonumber \\
 &&\mbox{} - 2\la N(\xv,t)P(m, \xv,t)  + 2 i \sqrt{\la} \xi (\xv , t)
P(m, \xv,t), \nonumber \\
 ~\label{ssm}
 \end{eqnarray}
 where $P*P= \int_{0}^{m} dm' P(m-m', \xv,t) P(m', \xv,t)$ is a
convolution term, $\xi (\xv,t)$ is Gaussian white noise and $N(\xv, t) =
\int_{0}^{\infty} dm \bar{P}(m , \xv , t)$, is the local particle density.  
We are interested in $\pb = \langle P(m, \xv, t) \rangle$, where $\langle
\ldots \rangle$ denotes averaging with respect to the noise $\xi$.
If the initial number of
particles of different masses at different lattice sites are independent
Poisson random variables parameterized by initial average mass
distribution ${P}_{0} (m)$, then the initial condition to be supplied
with Eq.~(\ref{ssm}) is $P(\xv, m, 0)= {P}_{0}(m)$. 
It is easy to check that Eq.~(\ref{ssm}) conserves
the average mass density $ \rho = \int m \bar{P}
(m) dm$. 

As particles aggregate the typical mass grows in time as $t^{d/2}$.
If we are interested in small masses, we need to consider
masses smaller than the typical mass, $m \ll \rho(Dt)^{d/2}$. This can
be achieved by considering 
$m/m_{0}$ to be fixed as $t \rightarrow \infty$,
where $m_0$ is the smallest mass at $t=0$. In this case,
the first term on the right hand side of Eq.~(\ref{ssm})  is almost
surely small compared to the other terms. Consequently, the small mass
behavior of the local mass distribution is described by the following
system of non-linear stochastic partial differential equations (SPDE): 
 \begin{eqnarray}
 \left( \frac{\partial}{\partial t} - D \Delta \right) P (m, \xv, t) &=& -
2\la N(\xv,t)P(m, \xv,t)\nonumber \\
 &+& 2i\sqrt{\la} \xi (\xv , t) P(m, \xv,t), \label{n} \\ \left(
\frac{\partial}{\partial t} - D \Delta \right) N (\xv, t)  &=& - \la
N^2(\xv,t) \nonumber \\
 &+& 2i\sqrt{\la} \xi (\xv , t)  N(\xv,t). \label{N}
 \end{eqnarray}

Equations~(\ref{n}) and (\ref{N}) demonstrate an interesting connection
between this model and $A+A\rightarrow A$ model. 
Stochastic field $P$ can be
identified with $\frac{\partial N}{\partial N_{0}}$, where $N_{0}$ is the
initial density. Differentiating Eq.~(\ref{N}) with respect to
$N_{0}$, and setting $P=\frac{\partial N}{\partial N_{0}}$, we obtain
Eq.~(\ref{n}). 

We are interested in the behavior of $\pb$ in the limit of fixed $m$
and $t \rightarrow \infty$. We can then identify the particles with
this fixed mass as $B$ kind of particles and the remaining particles as
$A$ kind of particles. Then, clearly, 
the study of Eqs.~(\ref{n}) and (\ref{N}) is equivalent to
the study of a two species reaction
 \begin{eqnarray}
 A+A &\stackrel{\la}{\rightarrow}& A , \nonumber\\
 A+B &\stackrel{\la}{\rightarrow}& \rm{Inert}, \label{ab}
 \end{eqnarray}
in the limit when concentration of $B$ particles
is much smaller than that of $A$ particles. 
This two species problem has been studied in
$d=1$ for arbitrary diffusion rates \cite{Fisher}. 
Specializing results of this paper to our case, we find that $\pb \sim
t^{-3/2}$ for $t \rightarrow \infty$. Assuming that the large time
asymptotics of $\pb$ is universal, we can restore the $m$-dependence 
using dimensional analysis, to
obtain
 \begin{equation}
 \pb = C\frac{m}{\rho t^{3/2}}, \label{fisher}
 \end{equation}
 where $C$ is a constant, and $\rho$ is the average mass density.
Equation~(\ref{fisher}) matches with the exact results obtained for the
CCA model in one dimension \cite{Spouge,satya}.

However, no exact solutions are available for dimensions $d >1$.
In the rest of the paper we will be analyzing Eqs.~(\ref{n}) and (\ref{N}) 
in $d >1$ using the dynamical renormalization group method. 
We will show that for small masses and $1\leq d <2$,
\begin{equation}
 \pb \sim \frac{1}{\rho (Dt)^d} \bigg( \frac{m}{\rho (Dt)^{d/2}} \bigg) ^{e_{KR}},~ m \ll
\rho (Dt)^{d/2}, \label{krd}
\end{equation}
 where $e_{KR}=\epsilon +O(\epsilon ^2)$ and $\epsilon =2-d$. If $d=2$,
 \begin{eqnarray}
 P(m,t) &\sim& \frac{1}{\rho (Dt)^2} \ln(t/t_0) 
\ln \bigg(\frac{m}{\rho D t_0} \bigg) \bigg(1+ \frac{1}{\ln(t/t_0)} \bigg), \nonumber  \\ 
&& \mbox{for}~~ \rho D t_0 \ll m \ll \rho Dt\ln(t/t_0). \label{kr2}
 \end{eqnarray}
Here, $t_0 \sim \Delta^2/D$, where $ \Delta$ is the lattice spacing.

Before doing the renormalization group analysis, let us
analyze Eqs.~(\ref{n}) and (\ref{N}) in the mean field 
(weak coupling) limit. Neglecting stochastic terms in the right hand side of
Eqs.~(\ref{n}) and
(\ref{N}) and solving the resulting system of ordinary differential
equations, we obtain
 \begin{eqnarray}
 \bar{N}_{MF}(t)&=&\frac{N_{0}}{1+N_{0}\lambda t}, \label{Ncl}\\
 \bar{P}_{MF}(t)&=&\frac{P_{0}}{(1+N_{0}\lambda t)^2} . \label{Pcl}
 \end{eqnarray}

Thus, at large times, $\bar{P}(t) \sim t^{-2}$, given that mean field
theory is applicable. Relative corrections to the mean field result
are of the
order $g_{0} (t)=\frac{\lambda t}{(Dt)^{d/2}}$. Therefore mean field
theory is asymptotically exact in $d>2$ \cite{footnote}
and diverges with
$t$ if $d<2$. Re-summation of the most divergent terms in the weak
coupling expansion of $\bar{P}$ around the classical solution is required
and can be performed in the case at hand using the formalism of
renormalization group. The details of the computation are given in the
next Sec.~\ref{sec3}. Here, we would like to demonstrate that accounting for
renormalization of the coupling constant alone doesn't yield the correct
decay exponent as mentioned earlier.
To the leading order in $\ep = 2-d$, renormalization of the
effective reaction rate reduces to replacing $\lambda$ in Eqs.~(\ref{n})
and (\ref{N}) with renormalized value $\lambda _{R} =f(\epsilon)
t^{-\ep/2}$ and omitting stochastic terms (renormalized mean field
approximation). Eliminating $\nb$ from the resulting system of ordinary
differential equations one
finds the following equation for $\pb$:
 \begin{equation}
 \frac{\partial \bar{P}}{\partial t} = -d \frac{\bar{P}}{t} . 
 \label{Pclr}
 \end{equation}
 This implies, that $\pb \sim m^{0}t^{-d}$.  In other words, $\pb$ scales
with time according to its physical dimension. As a result it does not
depend on mass. As we will show in Sec.~\ref{sec3},
arguments leading to this
conclusion are incorrect, as they disregard the possibility of anomalous
dimension of the stochastic field $P$.

\section{\label{sec3}Renormalization group analysis of stochastic aggregation.}

\begin{figure}
 \includegraphics[width=8.0cm]{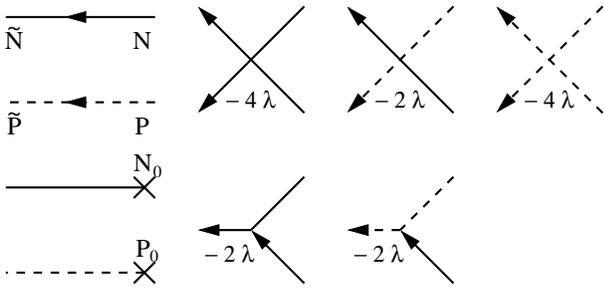}
 \caption{\label{fig1} Propagators and vertices of the effective theory,
Eq.~(\ref{act})}. 
 \end{figure}

The average mass distribution $\bar{P}(t)$ and average particle density $\nb$
admit functional integral representations which can be obtained
by applying the Martin-Siggia-Rose procedure \cite{MSR} to
Eqs.~(\ref{n}) and (\ref{N}) (equivalently see Eqs.~(2)-(4) of
\cite{Oleg}). We then obtain
\begin{eqnarray}
 \langle O (t) \rangle &=& \int \mathcal{D}N (\vec{x}',t') 
\mathcal{D}\tN(\vec{x}',t') \mathcal{D}P(\vec{x}', t')  \mathcal{D}\tP
(\vec{x}',t') \nonumber \\
 &&\mbox{}\times O(t) ~ e^{-S_{eff}[N, \tN, P, \tP]}, \label{fi}
\end{eqnarray}
 where $O(t)=N(\xv ,t)$ or $P(\xv ,t)$ and
 \begin{eqnarray}
 \lefteqn{S_{\mbox{eff}}=\int_{0}^{t} d^{d}x d\tau \big[ \tN (\dot{N} -D
\Delta N)  + \tP (\dot{P}-D\Delta P)} \nonumber  \\
 &+& \!\!\!\lambda (\tN N^2 \!\!+2 \tP P N 
 +\tN ^2 N^2 \!\!+2 \tN\tP N P +\tP ^2 P^2) \big] \nonumber \\ 
~\label{act}
 \end{eqnarray}
 is the effective action functional. Perturbative expansions of $\nb$,
$\pb$ in powers of $\lambda$ can now be obtained in the standard way, see for
example \cite{Drouffe}. Feynman rules for constructing terms of these
expansions are summarized in Fig.~\ref{fig1}. Due to the non-renormalization
of the diffusion rate as well as the average mass density
in the field theory Eq.~(~\ref{act}), in all that follows,
we use units in which $\rho=D =1$. 

The average mass distribution $\pb$ is formally given by the sum of all
diagrams built out of blocks presented in Fig.~\ref{fig1} with one outgoing
punctuated line ($P$-line). The contribution from each diagram is a
function of $\lambda N_{0} t$ and 'bare' dimensionless reaction rate
$g_{0} (t)= \frac{\lambda t}{t^{d/2}}$. Unless we are interested in the
small time expansion of $\pb$, $\lambda N_{0} t$ cannot be treated as a
small parameter. Therefore, the contributions of all diagrams proportional
to a given power of $g_{0} (t)$ and various powers of $\lambda N_{0} t$
have to be summed up. A simple combinatorial argument (see \cite{Lee} for
details) shows that the contribution of a diagram with $n$ loops is
proportional to $g_{0}(t)^n$. In the weak coupling regime the main
contribution to $\pb$ and $\nb$ is, therefore, given by the sum of all tree
diagrams, the first correction comes from summing all one-loop diagrams and so
on. It turns out \cite{Lee,Howard} that the sum of all tree
diagrams gives the mean field answers Eqs.~(\ref{Ncl}) and (\ref{Pcl}). The
expansion in powers of $g_{0} (t)$ is therefore the standard loop
expansion around solutions of mean field equations. It is obvious that in
$d<2$ the loop expansion is not very useful at large times as
$\lim_{t\rightarrow \infty}g_{0} (t) =\infty$. Fortunately, the value of
dimensionless reaction rate properly renormalized to account for the
build-up of inter particle correlations turns out to be of order
$\epsilon=2-d$ for large times. This allows one to convert the loop
expansion into an $\epsilon$-expansion using perturbative renormalization
group method. Such an expansion works well for large times and will
therefore yield all the information we need about the behavior of the
average mass distribution in the strongly fluctuating regime.
 \begin{figure}
 \includegraphics[width=8.0cm]{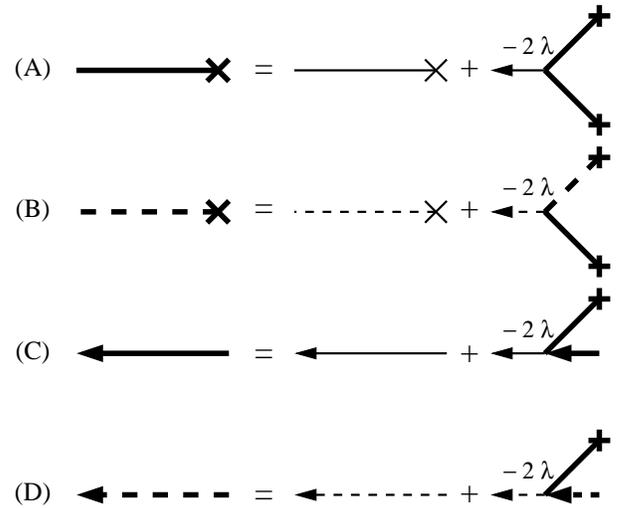}
 \caption{\label{fig2} Diagrammatic form of mean field equations for: (A) 
Average particle density; (B) Average mass distribution; (C) $NN$-response
function; (D) $PP$-response function.}
 \end{figure}

In computing loop corrections to any order, there are generally an infinite
number of diagrams to sum. However these diagrams 
can be resummed in part if one introduces classical response functions 
$G^{NN}_{cl}$
and $G^{PP}_{cl}$. Response function $G^{NN}_{cl}$ ($G^{PP}_{cl}$) is
equal to the sum of all tree diagrams with one outgoing and one ingoing
line of types N (P).  As was already mentioned, mean field densities
Eqs.~(\ref{Ncl}) and (\ref{Pcl}) are also equal to the sums of tree diagrams
with one outgoing N or P line correspondingly. 
One then simply has to associate 
incoming lines with mean field densities and
propagator lines with mean field response functions. Integral equations
satisfied by classical densities and response functions are presented in
diagrammatic form on Fig.~\ref{fig2}. 

Solutions of Eqs.~Fig.~\ref{fig2}(A) and Fig.~\ref{fig2}(B) coincide with 
Eqs.~(\ref{Ncl}) and (\ref{Pcl}), as they should. Equations 
Fig.~\ref{fig2}(C) and Fig.~\ref{fig2}(D) can also be solved with
the result
 \begin{eqnarray}
 \lefteqn{ G^{NN}_{cl} (x_{2}, t_{2};x_{1}, t_{1}) =G^{PP}_{cl}(x_{2},
t_{2};x_{1}, t_{1})=}\nonumber \qquad \qquad \\
 &&\bigg(\frac{N_{0}(t_{2})}{N_{0}(t_{1})}\bigg)^2 G_{0} (x_{2}-x_{1},
t_{2}-t_{1}), \label{prop}
 \end{eqnarray}
 where $G_{0}$ is the Green's function of the linear diffusion
equation.

\begin{figure}
 \includegraphics[width=8.0cm]{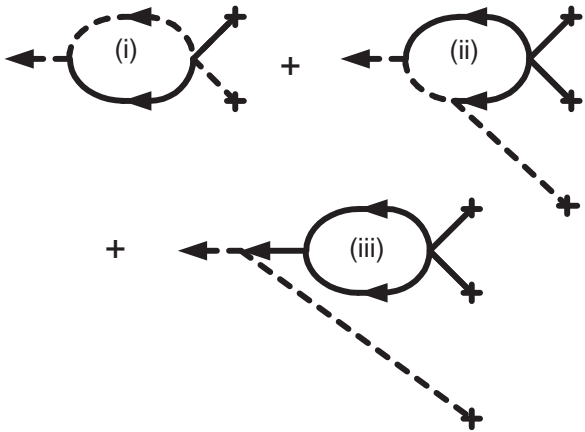}
 \caption{\label{fig3} One-loop corrections to the mean field answer for
average mass distribution.}
 \end{figure}

Using the notion of mean field response functions and densities one can
easily classify all one-loop diagrams contributing to average mass
distribution. The result is presented in Fig.~\ref{fig3}. 
A quick check shows that analytical expressions corresponding to diagrams
Fig.~\ref{fig3}(i) and Fig.~\ref{fig3}(iii) containing primitive loops 
diverge in $d\geq 2$, which is
consistent with that fact that upper critical dimension of the effective
theory Eq.~(\ref{act}) is two. Computing the relevant integrals in $d=2-\ep$
dimensions we find the following one-loop expression for the
average mass distribution:
\begin{eqnarray}
\lefteqn{\bar{P}(t)= \frac{P_{0}}{N_{0}^2} \left( \left( \frac{1}{\lambda
t}\right)^2+\frac{1}{(8\pi)^{d/2}} \frac{1}{\lambda t^{(2-\ep
/2)}} F(\ep ) \right)} \nonumber\\
&\times&\!\!\!\!\left(\!\! 1+O\!\!\left( \frac{1}{\lambda N_{0} t}\right)\!\!\right)
+ \mbox{2-loop corrections}, 
\label{Pt0}
\end{eqnarray}
 where $F(\ep)=\frac{4}{\ep}\frac{1-\ep/2}{(1+\ep/2)^2 (1+\ep/4)}$.  

Equation~(\ref{Pt0}) can be used
to determine the large time asymptotics of $\bar{P}(t)$ in
the following way. The exact average mass distribution
satisfies the Callan-Symanzik equation which we will
derive below. Coefficients of this equation depend on the law
of renormalization of all relevant couplings of the theory Eq.~(\ref{act}).
One relevant coupling is the effective reaction rate. Its  
renormalization is known ~\cite{Peliti}.
Below we will show that the only other relevant coupling is the initial
mass distribution $P_{0}$. We will determine it's renormalization law with
one-loop precision demanding that, the expression 
Eq.~(\ref{Pt0}) be non-singular in the limit
$\epsilon \rightarrow 0$ when expressed in terms of renormalized coupling
constant and renormalized initial mass distribution. This will 
determine coefficients of Callan-Symanzik equation up to terms of order
$g_{R}$, where $g_{R}$ is the renormalized dimensionless reaction rate.
Solving this equation we will be able to
compute the decay exponent of  
$\bar{P} (t)$ up to terms of order $\epsilon$.

Dimensional analysis of effective vertex parts of the theory Eq.~(\ref{act})
shows that the only relevant $bulk$ coupling is the effective reaction
rate. Its relevance is due to the recurrence 
of random walks in $d\leq 2$. Reaction rate
renormalization accounts for all fluctuation
effects in the $A+A\rightarrow A$ model. 

However, the CCA model is more complicated and interesting due to
the presence of $boundary$-relevant couplings. To identify them, 
we use the following version of
dimensional analysis. Boundary coupling constants correspond to vertices
with no incoming lines. As $PP$-interactions can be neglected in the
problem at hand, we are interested in boundary vertices
with at most one $P$-line. Assume for simplicity that $d=2$ (critical
dimension). Assume also that the initial density $N_{0}=\infty$. This
assumption is justified if one is interested in the large time behavior of
correlation functions in aggregation, as $N_{0}$ flows to infinity under
renormalization group transformation to increasingly larger time scales,
see \cite{Cardy1} for more details. Let $\Gamma_{\alpha, \beta} (t)$, where
$\alpha =0,1;~\beta=0,1,2,\ldots$ be the simultaneous Green's function of
the theory Eq.~(\ref{act}) with $\alpha$ $P$-lines and $\beta$ N-lines with
all external momenta equal to $0$. Using Eq.~(\ref{prop}), one can
express $G_{\alpha, \beta} (t) $ in terms of corresponding vertex part
$V_{\alpha, \beta} (t)$ as follows: 
 \begin{eqnarray}
 G_{\alpha, \beta}=\left( \frac{1}{t^{2}} \right) ^{\alpha+\beta}
\int_{0}^{t} d\tau \tau^{2 (\alpha+\beta)} V_{\alpha, \beta} (\tau).
\label{bt}
 \end{eqnarray}
 As physical dimension of $G_{\alpha, \beta} (t)$ is $(Length)^{-2
\beta-4 \alpha}$, $V_{\alpha, \beta}(t) \sim t^{-\beta-2
\alpha-1}$. As a result, Eq.~(\ref{bt}) converges at small times for any
$\alpha$ if $\beta>0$. If $\beta=0$ and $\alpha=1$, Eq.~(\ref{bt})  diverges
logarithmically. This divergence can be regularised using a small-time
cut-off $\frac{1}{\lambda N_{0}}$ and leads to the renormalization of the
initial mass distribution $P_{0}$. The latter plays a role of the coupling
associated with $V_{1,0}$. 

As a result of the discussed divergence, diagrams involving $V_{1,0}$ grow
with time faster compared with diagrams with the same number of loops but
with no sub-diagrams contributing to renormalization of $P_{0}$.
Consequently, diagrams belonging to the former class have to be re-summed
exactly in order to obtain the correct large time behavior of the average
mass distribution. 

We conclude that fluctuations in stochastic aggregation lead to two
effects: renormalization of the effective reaction rate $and$
renormalization of the initial mass distribution. It follows
that renormalization of these two couplings regularizes 
the perturbative expansion of $\bar{P}(m,t)$ to all orders.

As it turns out, the renormalization of $P_{0}$
is solely responsible for Kang-Redner anomaly.  In 
Sec.~\ref{sec5} we will analyze initial density renormalization by explicitly
re-summing small time divergences in the perturbative expansion for
$\pb$.  Now, we will show how it appears formally within the
framework of perturbative renormalization group method. 
We follow dynamical renormalization group procedure described in
\cite{Cardy1}, part I. For a review of results concerning coupling
constant renormalization, see ibid, part II.

Let us fix a reference time $t_{0}>0$. Expression (~\ref{Pt0}) evaluated
at $t_0$ is to be made finite by absorbing the divergences appearing as
$\ep\rightarrow 0$ into the renormalization of the reaction rate and the
initial density. 

Let $g_{0}=\lambda t_{0}^{\ep/2}$ be the dimensionless 'bare' reaction
rate. As has been shown in \cite{Peliti}, renormalized reaction rate
$g_{R}$ is related to $g_{0}$ via the following exact formula: 
 \begin{equation}
 g_{R} =\frac{g_{0}}{1+\frac{g_{0}}{g^{*}}}. \label{gr}
 \end{equation}
 Here $g^{*}= 2\pi \ep (1+O(\ep))$ is the non-trivial fixed point of the
renormalization group flow in the space of effective couplings of
Eq.~(\ref{act}). Recall that $g_{0} \sim \lambda$. Hence $\lim_{\lambda
\rightarrow \infty} g_{R} =g^{*}$. It follows from the
Callan-Symanzik equation that large time
behavior of the $\pb$ is also determined by the fixed point value of the
effective reaction rate $g^{*}$.  We therefore conclude that the limits
$t>0, \lambda \rightarrow \infty$ and $\lambda >0, t \rightarrow \infty$
of Kang-Redner model belong to the same universality class
as claimed in the introduction.

Solving Eq.~(\ref{gr}) with respect to $g_{0}$, expanding the result in
powers of $g_{R}$ and substituting the expansion into Eq.~(\ref{Pt0}) we
obtain the average mass distribution at time $t_{0}$ as a power series in
$g_{R}$: 
 \begin{eqnarray}
 \bar{P}(t_{0})=\frac{P_{0}}{N_0^2 g_{R}^2 t_{0}^d} \left[ 1-\frac{1}{g^{*}}
\big( 1+O(g^{*}) \big) g_{R} +O(g_{R}^2)\right]. 
 \label{Pint}
 \end{eqnarray}
As expected, the order-$g_{R}$ term in Eq.~(\ref{Pint})  is still
singular at $\ep=0$. To cancel the remaining divergences we have
to introduce renormalized initial mass distribution $P_{R}$: 
 \begin{equation}
 P_{R} = Z(g_{R}, t_{0}, \ep) P_{0}, \label{Z}
 \end{equation}
 where $Z$ is a power series in $g_{R}$ with coefficients chosen in such a
way, that the average mass density
 \begin{equation}
 \bar{P} (t, g_{R}, P_{R}, t_{0})=Z(g_{R}, t_{0}, \ep)  \bar{P}(t,
\lambda_{0}, P_{0}, \ep), \label{Pr}
 \end{equation}
 when expressed in terms of $P_{R}, g_{R}$, is non-singular at $\ep=0$. 
Substituting Eq.~(\ref{Pint}) into Eq.~(\ref{Pr}) we find that
 \begin{equation}
 Z=1+\frac{g_{R}}{g^{*}} +O(g_{R}^2), \label{Z1lp}
 \end{equation}
 in order for $\bar{P}(t)$ to be non-singular at 1-loop level. 

Now we can derive the Callan-Symanzik equation.
The fact that $
\bar{P}(t, \lambda_{0}, P_{0}, \ep)$ does not depend on the reference time
$t_{0}$ leads to the following equation for $\bar{P}(P_{R})$: 
 \begin{eqnarray*}
 t_{0}\frac{\partial}{\partial t_{0}} (Z^{-1} \bar{P}(P_{R}))=0. 
 \end{eqnarray*}
 Noticing that $\bar{P}(t) = t_{0}^{-d} \frac{P_{R}}{N_{0}^2}\Phi(t/t_{0},
g_{R})$, where $\Phi$ is a dimensionless function, one can convert the
above condition into Callan-Symanzik equation for $\bar{P} (t)$: 
 \begin{equation}
 \bigg( t\frac{\partial}{\partial t} + \frac{1}{2}\beta(g_{R}) 
\frac{\partial}{\partial g_{R}} + d-\frac{1}{2} \gamma (g_{R})\bigg)
\bar{P}(t, g_{R}, P_{R}, t_{0}) =0,\label{CS}
 \end{equation}
 where $\beta (g_{R}) = -2\frac{\partial}{\partial t_{0}} g_{R}$ is the
beta function of the theory Eq.~(\ref{act}) and 
 \begin{equation}
 \gamma (g_{R})  = -2
\frac{1}{Z} t_{0}\frac{\partial}{\partial t_{0}} Z = 
- \frac{g_R}{2 \pi} + O(g_{R}^2, g_R ~\ep),
\label{gamma}
 \end{equation}
 is the gamma function. 

It is well known, that at large times and in $d<d_{c}=2$, solutions to
Eq.~(\ref{CS}) are governed by non-trivial fixed points (zeroes) of
beta-function. Differentiating Eq.~(\ref{gr}) with respect to $t_{0}$, one
finds that $\beta (g_{R})=g_{R} (g_{R}-g^{*})$. Hence, $\beta
(g_{R})$ has a unique non-trivial fixed point $g_{R}=g^{*}$. It follows
from Callan-Symanzik equation (Eq.~(\ref{CS})) that $\bar{P}(t) \sim t^{-d^{*}},
~t\rightarrow \infty, $ where
 \begin{equation}
 d^{*} = d - \frac{1}{2} \gamma (g^{*}).\label{d}
 \end{equation}
 We see that scaling dimension of $\pb$ differs from it's physical
dimension by a term proportional to the value of $\gamma$-function at the
fixed point. This term is called the anomalous dimension of the field $P$.
The physical reason for the presence of anomalous dimension is the
renormalization of the initial mass distribution by small time
fluctuations. 

As $g^{*}\sim \ep$, the substitution of $g_{R}$ expansion of $\gamma$ into
Eq.~(\ref{d}) yields the $\ep$-expansion for $d^{*}$.  Using Eq.~(\ref{gamma}) 
we find that
 \begin{equation}
 d^{*} = d + \frac{1}{2}\ep + O(\ep^2).\label{de}
 \end{equation}
 Finally, one can restore the $m$-dependence of $\pb$ using dimensional
argument. The result is
 \begin{equation}
 \pb = A(\ep) \frac{1}{t^{d}} \bigg( \frac{m^{\frac{1}{d}}}{\sqrt{t}}
\bigg)^{e_{KR}}, \label{pmt}
 \end{equation}
 where $e_{KR}=\ep+O(\ep ^2)$ is equal to twice the anomalous dimension of
the stochastic field $P$. 

Note that in $d=1$, $e_{KR}=1$ and $\pb \sim m^{1}$, which coincides
with the exact answer, \cite{Spouge}. However, at the moment we do not
have reasons to believe, that higher order $\ep$-corrections to our answer
for $e_{KR}$ vanish identically for any $0 < \ep \leq 1$. 
Equation~(\ref{de}) implies that anomalous dimension of $P$ vanishes at 
$d_{c}=2$.
Yet, it follows from general theory of perturbative renormalization group
that traces of anomalous dependence of $P(m,t)$ on mass in $d<2$ must be
present at the critical dimension as well. We analyze Kang-Redner
anomaly in $d=2$ in Sec.~\ref{sec4} and compare renormalization group
predictions with conclusions of direct numerical simulations of the system
of diffusing-aggregating point particles on the two-dimensional lattice. 

\section{\label{sec4}Kang-Redner Anomaly in Two Dimensions.}

It is well known that anomalies in $d_{c}-\epsilon$ dimensions lead to
logarithmic corrections to mean field theory answers in $d=d_{c}$.
Kang-Redner anomaly is not an exception. Small mass behavior of the
average mass distribution in two dimensions can be easily obtained by
solving the Callan-Symanzik equation (see Eq.~(\ref{CS})).  Note that
 \begin{eqnarray}
 \beta (g) \mid _{d=2}&=& \frac{1}{2\pi} g^2, \nonumber\\
 \gamma (g) \mid _{d=2}&=&-\frac{1}{2\pi} g (1+O(g^2)). \label{gb}
 \end{eqnarray}
 Solving Eq.~(\ref{CS}) with coefficients given by Eq.~(\ref{gb}) and the 
initial condition
 \begin{equation}
 \bar{P} (t_{0})=\rm{Const}  \frac{1}{g_{R}^2 t_{0}^2}
(1+O(g_{R})), \label{ic}
 \end{equation}
 produced by the mean field theory, we find that
 \begin{equation}
 \bar{P} (t) = \rm{Const}  \frac{\ln(t/t_{0})}{t^2}
(1+O(1/\ln(t/t_{0})). \label{twod}
 \end{equation}
 Such a behavior of $\pb \mid _{m=fixed}$ in two dimensions was originally
seen by Kang and Redner in numerical simulations \cite{KR}. We
have now shown that Eq.~(\ref{twod}) can be obtained as a result of
systematic renormalization group computation. 
 \begin{figure}
 \includegraphics[width=8.0cm]{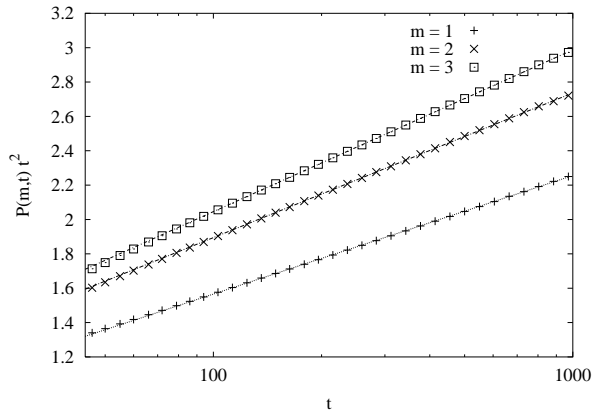}
 \caption{\label{fig5}The variation of $\bar{P}(m,t) t^2$ with $t$ is shown for
$m=1,2$ and $4$ on a semi-log scale. The variation is linear implying that
$\bar{P}(m,t) = c(m) \ln(t)/t^2$, where $c(m)$ is some mass dependent
function.}
 \end{figure}
 \begin{figure}
 \includegraphics[width=8.0cm]{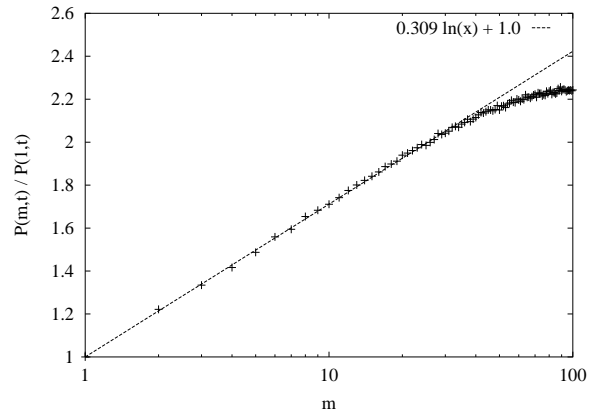}
 \caption{\label{fig6}The small $m$ behavior of $\bar{P}(m,t)/\bar{P}(1,t)$ 
is shown on a semi-log plot. The variation is linear implying that $c(m)
\sim \ln(m)$ with $c(m)$ as defined in caption of Fig.~\ref{fig5}.}
 \end{figure}

Analyzing their model in two dimensions, Kang and Redner noticed that
$\pb$ is a slowly varying (increasing, as suggested by Fig.~\ref{fig2} of
\cite{KR}) function of mass. We can now quantify this observation by
computing the mass dependence of the average mass distribution. In
$d=d_{c}$ direct dimensional arguments cannot be used to restore the mass
dependence of Eq.~(\ref{twod}) due to the explicit dependence of $\pb$ on the
lattice cut-off. Instead let us analyze Eq.~(\ref{pmt}) for
$\bar{P}(m,t)$ in the limit $\epsilon \rightarrow +0$. Near $\ep=0$, the
amplitude $A(\epsilon) \sim \ep^{-2}$. Expanding the right hand side of
Eq.~(\ref{pmt}) in powers of $\ep$ and using the fact that $\bar{P}(m,t;\ep)$
is non-singular at $\ep=0$ as a function of renormalized parameters, we
find that
 \begin{eqnarray}
 \pb = \frac{1}{t^2} \bigg( C_{1} \big( \ln(t/t_{0}) \big)^2+C_{2}
\ln(t/t_{0})\ln(m/t_{0})\nonumber\\
 + C_{3}\big( \ln(m/t_{0} \big))^2\bigg) \bigg( 1+O \big(
1/\ln(t/t_{0})\big) \bigg). \label{step}
 \end{eqnarray}
 We are interested in $\pb$ in the limit $t \rightarrow \infty,
~m=\mbox{fixed}$. Time dependence of $\pb$ is given by Eq.~(\ref{twod}). 
Therefore, coefficient $C_{1}$ in Eq.~(\ref{step}) is $0$. Hence
 \begin{equation}
 \bar{P} (m,t) =\rm{Const} \frac{\ln(t/t_{0})  \ln(m/t_{0})}{t^2} \bigg(
1+O\big( 1/\ln(t/t_{0}) \big) \bigg). \label{mtwod}
 \end{equation}
 Note that Eq.~(\ref{mtwod}) is valid for $m\ll M(t)$, where $M(t)\sim
t/\ln(t/t_{0})$ is the typical mass. If $m \sim M(t)$, then $\pb \sim
(\ln(t/t_{0})/t)^2$, which coincides with the answer for $\pb
\mid_{m\sim M(t)}$ in the intermediate mass range \cite{Oleg}. 

To check our analytical results Eq.~(\ref{twod}) and Eq.~(\ref{mtwod}) we
did a numerical simulation of the system on a two-dimensional lattice. For
the sake of computational effeciency, 
we chose to work in the limit $\la \rightarrow
\infty$ so that a lattice site can hold utmost one particle. The lattice
was chosen to be a square lattice of size $3000 \times 3000$ with
periodic boundary conditions. In Fig.~\ref{fig5}, we show the variation
of $\bar{P}(m,t)$ with $t$ for fixed small $m$. It is seen that
$\bar{P}(m,t) = c(m) \ln(t)/t^2$, where $c(m)$ is some function of the
mass $m$. This is in excellent agreement with Eq.~(\ref{twod}). To
determine $c(m)$, we studied the variation of
$\bar{P}(m,t)/\bar{P}(1,t)$ with the mass $m$ (see Fig.~\ref{fig6}). We
see that $c(m) \sim \ln(m)$, thus confirming Eq.~(\ref{mtwod}).

\section{\label{sec5}Kang-Redner anomaly via explicit resummation of boundary
divergences}

In this section we will derive Eq.~(\ref{pmt}) without using the
formalism of renormalization group. Instead, we will identify the
principal set of diagrams contributing to the the large time limit of
$\pb$ and derive a simple integral equation satisfied by the sum of these
diagrams. 

Additional divergences in the terms of the perturbative expansion of the
average mass distribution discussed in the previous Sec.~\ref{sec4}
are actually
due to the quadratic singularity of the mean field expression Eq.~(\ref{Pcl})
for the average mass distribution at $N_{0}=\infty ,~ t=0 $. Let us
illustrate this statement on the example of diagram (i) of Fig.~\ref{fig3}.
Assuming for simplicity that $d=2$ and denoting ultraviolet cut-off by
$\Delta t$, we find that the most divergent contribution coming from this
diagram is
 \begin{equation}
 I_{(i)} \sim \frac{P_{0}}{2\pi \lambda^2 N_{0}^2 t^2} \ln \bigg(\frac{t}{\Delta
t} \bigg)\int_{0}^{t} \!\!\! d\tau \frac{\lambda N_{0}}{1+\lambda N_{0} \tau}
(1+O(1/\lambda N_{0} t)). \label{div}
 \end{equation}
 Setting $N_{0}=\infty$ under the sign of integration is clearly
impossible, as the resulting integral would 
diverge. Direct computation of
Eq.~(\ref{div}) shows that
 \begin{equation}
 I_{(i)} \sim \frac{P_{0}}{2\pi \lambda^2 N_{0}^2 t^2} \ln \bigg(\frac{t}{\Delta
t} \bigg) \ln(\lambda N_{0} t) (1+O(1/\lambda N_{0} t)). \label{bndr}
 \end{equation} We see that $I_{(i)}$ is proportional to $\ln(t)^2$, not
$\ln(t)$ as one would have deduced from the simple loop counting. The
additional singularity is due to the fact that $\bar{P}_{MF}\mid
_{N_{0}=\infty} \sim t^{-2}$. (Recall that $\bar{N}_{MF}\mid
_{N_{0}=\infty} \sim t^{-1}$. As a result divergences at $t=0$ are absent
in the field theory of $A+A\rightarrow 0$ reaction.)  The additional
$boundary$ singular terms are generally present in all Feynman integrals
contributing to $\pb$ and have to be re-summed to all orders of
perturbation theory to get the correct large time asymptotics of the
average mass distribution. 

\begin{figure}
 \includegraphics[width=8.0cm]{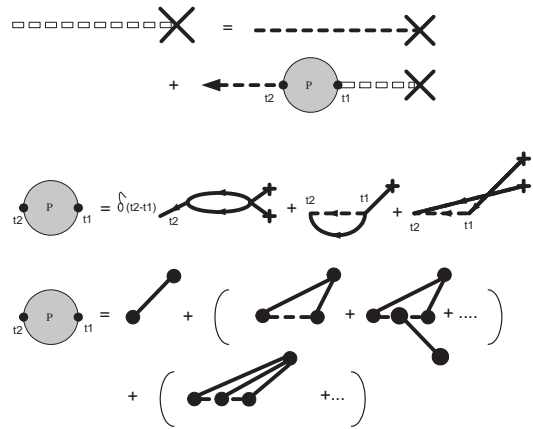}
 \caption{\label{fig4} (I) Schwinger-Dyson Equation for $\Pi (t)$. (II)
Perturbative expansion of the polarization operator:  (IIA)- in the number
of loops; (IIB)- in the order of exact vertex parts.}
 \end{figure}

Let $\Pi (t_{2}, t_{1})$ be the exact zero momentum vertex function of
type $\tP P$ - the sum of all one-particle irreducible diagrams
contributing to $\tP P$ response function divided by the propagators
corresponding to external lines. The Schwinger-Dyson equation for the
exact average mass distribution reads: 
 \begin{equation}
 \bar{P}(t)=\bar{P}_{MF} (t) +\frac{1}{t^2}\int_{0}^{t} dt_{2}
\int_{0}^{t_{2}} dt_{1} \Pi(t_{2}, t_{1}) \bar{P} (t_{1}). 
 \label{sd}
 \end{equation}
Equation~(\ref{sd}) is most easily derived using the formalism of Feynman
diagrams. Its diagrammatic representation is given in Fig.~\ref{fig4} (I). 

We know that in two dimensions $\bar{P} (t) \sim t^{-2}$.  Therefore, near
$d=2$, the function $\eta (t) = \bar{P}(t) \lambda t^2$ is slowly varying.
The following simplified version of Eq.~(\ref{sd}) is therefore valid
to the leading order in $\ep$: 
 \begin{equation}
 \eta (t)=1 +\int_{0}^{t} dt_{1} \Pi( t_{1}) \eta(t_{1}),
 \label{rsd}
 \end{equation}
 where
 \begin{equation}
 \Pi (t)= t^2 \int_{0}^{t}\frac{dt_{1}}{t_{1}^2} \Pi(t,t_{1}). 
 \label{pol}
 \end{equation}
 Differentiating Eq.~(\ref{rsd}) with respect to time, we find that $\eta (t)$
satisfies the following differential equation: 
 \begin{equation}
 \frac{d \eta}{dt} (t) =\Pi (t) \eta (t).\label{dif}
 \end{equation}
 It turns out (see below), that at large times
 \begin{equation}
 \Pi (t) = \frac{1}{t}\left( \frac{\ep}{2} +O(\ep ^2) \right)  \left(
1+O(1/t^{\ep/2}) \right). \label{pole}
 \end{equation}
 Solving Eq.~(\ref{dif}) with $\Pi (t)$ given by Eq.~(\ref{pole}), we
find that $\bar{P}(t) \equiv \eta (t)/t^2 \sim t^{(-2+\ep/2+O(\ep^2))}$.
Thus, we have been able to rederive Eq.~(\ref{de}) without any reference to 
renormalization group. 

Before we turn to the derivation of Eq.~(\ref{pole}), we would like to
demonstrate, that solving integral equation Eq.~(\ref{rsd}) is indeed
equivalent to summing leading boundary singular terms in the perturbative
expansion of $\bar{P}(t)$ to all orders. All divergences at $t=0,
N_{0}=\infty$ can be regularised by setting the lower limit of integration
in the integral in the right hand side of Eq.~(\ref{rsd}) equal to 
$\frac{1}{\lambda N_{0}}$. Resulting equation,
 \begin{equation}
 \eta(t)=1+\frac{\ep}{2} \int_{\frac{1}{\lambda N_{0} t}}^{t}
\frac{dt_{1}}{t_{1}} \eta(t_{1}),\label{int}
 \end{equation}
 can be solved using the method of consecutive approximations.  Setting
$\eta (t)= \eta_{0} (t)+\eta_{1}(t)+\eta_{2}(t)+\ldots$ and treating the
integral in the right hand side of Eq.~(\ref{int})  perturbatively, we find that
$\eta_{0}(t)=1$ and
 \begin{eqnarray}
 \eta_{n} (t)=\frac{1}{n!}(\frac{\ep}{2} \ln(\lambda N_{0} t))^n, \quad
n=1,2,\ldots
 \end{eqnarray}
 Note that $\eta_{1} (t)$ is equal to 1-loop boundary singular term,
$\eta_{2} (t)$ - 2-loop boundary singular term and so on.  Summing the
resulting series for $\eta$ one confirms our main result that $\eta (t)
\sim t^{\ep/2}$. One loop-boundary singularity leads to mass-dependent
logarithmic corrections to the average mass distribution. This is most
easily seen by expanding Eq.~(\ref{pmt}) in powers of $\epsilon$.  These
corrections, which were originally observed in \cite{Oleg}, motivated to a
certain extent the present investigation. 

The knowledge of the first term in the $\epsilon$-expansion of decay
exponent of $\eta (t)$ leads to a good approximation of the average mass
distribution , given that $\ep \ln(\lambda N_{0} t)  \sim 1$, but $\ep^2
\ln(\lambda N_{0} t) \ll 1$. If the latter condition breaks down, 
we need to
know the expansion of the  polarization operator $\Pi(t)$ 
to second order in
$\epsilon$ $and$ modify Eq.~(\ref{rsd}) by including corrections
proportional to the derivative $\eta (t)$ (which account for the fact that
$\eta (t)$ is not constant). 

We will now derive Eq.~(\ref{pole}) by computing $\Pi (t)$ to the first order
in $\ep$ using 1-loop diagrams and verifying that higher-loop diagrams
lead to terms of order $\ep^2$ and higher. 
If mean field theory for the average mass distribution was exact,
operator $\Pi$ would have been identically equal to zero.  Consequently
only loop diagrams contribute to $\Pi(t_{2}, t_{1})$.  The one-loop
diagrams are shown in Fig.~\ref{fig4} (IIA). The computation of corresponding
Feynman integrals is straightforward. After integration with respect to
$t_{1}$ we find their respective contributions to $\Pi (t)$: 
 \begin{eqnarray}
 \Pi_{(i)} (t) &=& \frac{- 8\lambda t^{\ep/2}}{(8\pi)^{d/2} t}
\frac{(1+2\ep)}{\ep (1+\ep/2)^2}, \label{pii}\\
 \Pi_{(ii)} (t) &=&\frac{8\lambda t^{\ep/2}}{(8\pi)^{d/2} t} \frac{1}{\ep
(1+\ep/2)}, \label{piii}\\
 \Pi_{(iii)} (t) &=& \frac{- 8 \lambda t^{\ep/2}}{(8\pi)^{d/2}
t}\frac{1}{(1+\ep/2)(2+\ep/2)}.  \label{piiii}
 \end{eqnarray}

Note that individual contribution from diagrams (i) and (ii) are
$1/\ep$-times bigger than the contribution from diagram (iii), which does
not contain primitive loops. Yet, terms of order $\ep ^{-1}$ cancel upon
adding $\Pi_{(i)} (t)$ and $\Pi_{(ii)}(t)$ leaving terms of order up to
$\ep^{0}$ - same as the leading order of terms of $\Pi _{(iii)}(t)$. This
cancellation explains why we had to account for an apparently subleading
contribution of diagram (iii) to the perturbative expansion of $\Pi(t)$.
Such a cancellation is not accidental and happens at all orders of loop
expansion: diagrams (i) and (ii) can be interpreted as first two terms in
$\ep$-expansion of the first term in the cumulant expansion of $\Pi(t_{2},
t_{1})$, see Fig.~\ref{fig4} (IIB). This term corresponds to an exact total
particle density $\bar{N} (t)$ connected to the $\tP P$-propagator via an
exact $\tP NP$-vertex.  The vertex is of the order of the fixed point
coupling $g^{*} \sim \ep$, while exact density is of the order $\ep
^{-1}$, \cite{Lee}.  Therefore the contribution of the term in question to
$\Pi (t_{2}, t_{1})$ is of order 1, which is reflected in cancellation of
the terms of the lowest order in $\ep$ in every term of its loop
expansion. 

Adding together Eqs.~(\ref{pii}), (\ref{piii}) and (\ref{piiii}), we find that
 \begin{equation}
 \Pi(t) = \frac{2\lambda t^{\ep/2}}{(8\pi)^{d/2}t}(1+O(\ep))+\mbox{2-loop
corrections}. 
 \label{pi0}
 \end{equation}

Large-t behavior of $\Pi (t)$ can be obtained from Eq.~(\ref{pi0}) by
replacing the 'bare' reaction rate $\lambda$ with renormalized reaction
rate $\lambda _{R} (t) \sim 2\pi \ep t^{-\ep/2}, t \rightarrow \infty$.
The result does indeed coincide with Eq.~(\ref{pole}). 

It remains to verify that two- and higher loop diagrams contribute only to
higher order terms in the $\ep$-expansion of $\Pi (t)$. Order-$\ep^{-2}$
contributions from diagrams containing only primitive loops cancel as
explained above. Order-$\ep^{-1}$ contributions from diagrams with two
primitive loops and two-loop diagrams with one primitive loop are
accounted for by one-loop renormalization of coupling constant in one-loop
diagrams. Hence non-trivial corrections to polarization operator come only
from two-loops diagrams containing no primitive loops and non-singular
parts of all other two-loop diagrams. Simple counting shows that
contribution from such diagrams to $\Pi(t)$ is proportional to
$\frac{(\lambda _{R} (t) t^{\ep/2})^n}{\ep^{(n-2)} t} \sim
\frac{\ep^2}{t}$. Similar argument shows that $n$-loop diagrams contribute
to $\Pi (t)$ at the order $\ep^n$ only. 

Now it is very easy to characterize the class of diagrams giving the
leading contribution to the $\ep$-expansion of $\Pi (t)$. The statistics
of $N(t)$ is strongly non-Gaussian. Yet, the main contribution to the
polarization operator $\Pi(t)$ comes from the diagrams proportional to the
first and the second cumulants of the stochastic field $N(x,t)$ only.
Non-Gaussian effects are due to the fact that these cumulants are
connected to $\bar{P}P$-propagator via exact vertices. 

It is also possible to derive formula Eq.~(\ref{twod}) without using the 
formalism
of renormalization group. Instead, one can solve equation Eq.~(\ref{dif})
directly using one-loop expression for polarization operator $\Pi (t)$ in
two dimensions. The latter can be obtained by setting in Eq.~(\ref{pi0}) $d=2$
and replacing the bare reaction rate $\lambda$ with renormalized reaction
rate in two dimensions: $\lambda _{R} (t) = \frac{4\pi}{\ln(t/t_{0})}
(1+O(1/\ln(t/t_{0})))$. The presented expression for $\lambda _{R} (t)$ is
easy to compute by explicit re-summation of all diagrams contributing to
renormalization of the bare reaction rate, see \cite{Lee} for details. The
resulting equation for $\eta (t)  \equiv t^2 P(t)$ is
 \begin{equation}
 \dot{\eta}(t) = \frac{1}{t \ln(t/t_{0})} \eta (t).
 \label{sd2d}
 \end{equation}
 The solution is $\eta (t) \sim \ln(t/t_{0})$. Correspondingly, $P(t) \sim
\frac{\ln(t/t_{0})}{t^2}$, which coincides with Eq.~(\ref{twod}).

\section{\label{sec6}Kang-Redner anomaly and corrections to Kolmogorov particle
spectrum.}

Now we will show that the mean field result $\pb \sim m^{0}$ can be
interpreted as a constant flux (Kolmogorov) solution of the Smoluchowski
equation. We will then interpret Kang-Redner anomaly as a breakdown of
Kolmogorov scaling due to strong flux fluctuations developing at large
times. 

Averaging Eqs.~(\ref{n}) and (\ref{N}) with respect to noise one gets the
following relation between one- and two point mass distribution functions: 
 \begin{eqnarray}
 \frac{\partial \langle P\rangle}{\partial t} &=& -\la \langle PN\rangle,
\label{asP} \\
 \frac{\partial \langle N\rangle}{\partial t} &=& -\frac{1}{2}\la \langle
N^2\rangle.
 \label{asm}
 \end{eqnarray}
Let us look for solutions of Eq.~(\ref{asm}) having the form
 \begin{equation}
 \pb=\frac{\bar{N}(t)}{M(t)} J(\mu, t), \label{p}
 \end{equation}
 where $\mu=\frac{m}{M(t)}$. Recall that $M(t)= \bar{N}^{-1}$ is the
typical mass. The new dependent variable $J$ has a simple physical
meaning:$\int_{0}^{\mu}d\mu ' J(\mu ', t)$ is the average number of
particles with masses less than $M(t)\mu$ contained in the volume
$\bar{N}^{-1}(t)$. 

At times much less than $t_{c}=(\la)^{-2/\ep}$ relative fluctuations of
local density are small. As a result, mean field theory is applicable and
$\langle JN \rangle \approx \langle J \rangle \langle N \rangle$. As a
result, equation Eq.~(\ref{asP})  simplifies to
 \begin{equation}
 \frac{\partial }{\partial t} (\frac{J}{\mu}) = \frac{\partial J}{\partial
\mu}. \label{flux}
 \end{equation}
 Therefore, $\frac{J}{\mu}$ is a locally conserved quantity with flux
equal to $(-J)$. Note that $J>0$. Therefore the cascade of $\frac{J}{\mu}$
is $inverse$ in the terminology of turbulence: its flux is directed
towards the small masses. We see that self-similar solutions of
Eq.~(\ref{asm}) correspond to constant flux solutions of Eq.~(\ref{flux}). The
latter is just $J=Const$. Constant flux solutions of kinetic equations are
called Kolmogorov solutions in the theory of weak turbulence, see
\cite{Zakh} for details. We therefore conclude that in the mean field
approximation
 \begin{equation}
 \langle P \rangle(m,t)=\frac{\bar{N}(t)}{M(t)} \mu^{e_{kolm}},
 \label{kolm}
 \end{equation}
 where $e_{kolm}=0$ is the exponent, which determines the Kolmogorov
scaling of the average mass distribution. Note, that the flux $J$ also has
a meaning of dimensionless particle density (the number of particles in
the volume $\bar{N}^{-1}$ , which is an obvious integral of motion. 

The fact that $J=const$, means that particles are equipartitioned between
system's degrees of freedom and particles' flux is identically equal to
zero. The characteristic feature of the state Eq.~(\ref{kolm}) of our system
is therefore the presence of non-zero constant flux of one integral of
motion and equipartition of the other. Similar kind of behavior has been
observed in models of turbulent advection, \cite{Nazar}

We know however, that mean field approximation is invalid in the limit of
large times if dimension is two or less because of strong fluctuations of
local particle density . Using results of the previous sections one can
interpret Kang-Redner anomaly as the anomaly in the constant flux
condition: 
 \begin{equation}
 \mu \frac{\partial J}{\partial \mu} =\frac{e_{KR}}{d} J,
 \label{anom}
 \end{equation}
 where $e_{KR}$ is Kang-Redner's exponent. 

Solving Eq.~(\ref{anom}), we find that $P(m,t) \sim \mu^{e}$, where
$e=e_{kolm} +\frac{e_{KR}}{d}$. Therefore, Kang-Redner anomaly can be also
interpreted as a correction to Kolmogorov scaling of the average mass
distribution due to strong fluctuations.

\section{\label{sec7}Conclusion.} 

In the present paper we have shown that the problem of cluster-cluster
aggregation in $d\leq 2$ can be effectively analysed using renormalization
group method. We have demonstrated that the dependence of average mass
distribution on mass is determined by the anomalous dimension of
the stochastic field $P$ (local mass distribution).  This anomaly
is due to the relevance of 'boundary'
($t=0$)  fluctuations for the large times asymptotics of $\pb$. In that
respect the phenomenon of Kang-Redner anomaly resembles the phenomenon of
boundary phase transition in equilibrium statistical mechanics, see
\cite{Cardy} for a review.  Formally, the anomalous dimension of the local
mass distribution is a consequence of the non-triviality of the
$\gamma$-function of the effective field theory Eq.~(\ref{act}). The fact that
$\gamma (g) \neq 0$ is ultimately responsible for the breakdown of the
Smoluchowski theory (or equivalently, the renormalized mean field theory) 
applied
to the model at hand. At present, renormalization group analysis seems to
be the only theoretical method of studying the problem of cluster-cluster
aggregation in $d>1$. This is not quite satisfactory, as the
analysis is essentially perturbative in nature.
However, our theoretical predictions concerning the behavior
of the average mass distributions at small masses in two dimensions
have been unambiguously confirmed numerically.

In section ~\ref{sec6} we have shown that there is a relation between
cluster-cluster aggregation and the theory of weak turbulence. In
particular, we demonstrated that Kang-Redner anomaly can be interpreted as
a correction to the Kolmogorov spectrum of particles in the mass space. 
However, we must stress that the analogy between our model and the
phenomenon of turbulence must be taken with a pinch of salt: while the
cascade of the conserved quantity in our model happens along the mass
axis only, conserved quantities (such as energy or enstrophy) in more
traditional turbulent systems flow through the scales of the physical
space.

The method of dynamical renormalization group developed in the context of
the model at hand can be applied to other non-equilibrium particle systems
as well. In particular the problem of cluster-cluster aggregation with
annihilation of particles,  
can be solved using techniques similar to those outlined 
in this paper \cite{Raj1}. This latter problem is related to
the computation of the domain wall persistence exponent for the
1d q-state Potts model \cite{Raj1}.

\section{Acknowledgments.}

The authors would like to thank John Cardy for many insights
into the applications of renormalization group.
Useful discussions with Robin Ball, Colm Connaughton, Igor
Kolokolov, Anttii Kupiainen, Vladimir Lebedev, Paolo Muratore-Ginanneschi,
Sergey Nazarenko are also appreciated. OZ would like to acknowledge
the hospitality of the Department
of Mathematics of Helsinki University where part of the present work was
done. The work at Oxford was supported by EPSRC, UK.


\begin{thebibliography}{1}
\bibitem{footnote} The statement of correctness of mean field theory in 
$d>2$ has to be understood properly. For instance, the effective reaction rate
entering the Smoluchowski coagulation equation is not equal to the bare 
reaction rate due to strong ultraviolet divergence of Feynman integrals
contributing to renormalization of the reaction rate. As a result,
density-density correlation function is not equal to the product of
one-point function, but rather is proportional to it. Yet, as the
renormalized reaction rate in $d>2$ is a constant depending on lattice
spacing, decay exponents are predicted by the mean field correctly.
\end{thebibliography}
\end{document}